\documentclass[aps,prb,reprint,showpacs,superscriptaddress,amsmath,amssymb]{revtex4-1}

\usepackage[utf8]{inputenc}

\usepackage{graphicx}
\DeclareGraphicsExtensions{.eps}

\usepackage{bm}
\usepackage{textcomp} 
\usepackage{enumitem}
\usepackage{amsmath,amssymb}
\usepackage{graphicx}
\usepackage{color}
\usepackage{xcolor}
\usepackage{amsmath}
\usepackage{qcircuit}
\usepackage{blindtext, rotating}
\usepackage{braket}
\usepackage{multirow}
\setlist{noitemsep,leftmargin=*}

\usepackage{hyperref}
\hypersetup{
	colorlinks=true,
	urlcolor= blue,
	citecolor=blue,
	linkcolor= blue,
	bookmarks=true,
	bookmarksopen=false,
}

\begin{document}

\title{Electronic and magnetic properties of VOCl/FeOCl antiferromagnetic heterobilayers}

\author{F. Mahrouche}
\affiliation{Laboratoire de Physique Th\'eorique, Facult\'e des Sciences Exactes, Universit\'e de Bejaia, 06000 Bejaia, Alg\'erie}
\affiliation{Departamento de F\'{i}sica Aplicada, Universidad de Alicante, 03690 San Vicente del Raspeig, Spain}

\author{K. Rezouali}
\affiliation{Laboratoire de Physique Th\'eorique, Facult\'e des Sciences Exactes, Universit\'e de Bejaia, 06000 Bejaia, Alg\'erie}

\author{Z. C. Wang}
\affiliation{QuantaLab, International Iberian Nanotechnology Laboratory (INL), 4715-330 Braga, Portugal}

\author{J. Fern\'{a}ndez-Rossier}
\affiliation{QuantaLab, International Iberian Nanotechnology Laboratory (INL), 4715-330 Braga, Portugal}
\thanks{On leave from Departamento de F\'{i}sica Aplicada, Universidad de Alicante, 03690 San Vicente del Raspeig, Spain.}

\author{A. Molina-S\'{a}nchez}
\affiliation{Institute of Materials Science (ICMUV), University of Valencia, Catedr\'{a}tico Beltr\'{a}n 2, E-46980, Valencia, Spain}

\begin{abstract}
We study the electronic properties of the heterobilayer of vanadium and iron
oxychlorides, VOCl and FeOCl, two layered air 
stable van der Waals insulating oxides with different types of antiferromagnetic
order in bulk: VOCl monolayers are ferromagnetic (FM)  whereas the FeOCl monolayers are antiferromagnetic (AF). We use density functional theory (DFT) calculations,  with  Hubbard correction that is found to be needed 
to describe correctly the insulating nature of these compounds. We compute the magnetic anisotropy and propose a spin model Hamiltonian. 
 Our calculations show that  interlayer coupling in weak and ferromagnetic so that  magnetic order of the monolayers is preserved in the  heterobilayers  providing thereby a van der Waals heterostructure that combines two monolayers with different 
magnetic order.  Interlayer exchange should lead both to exchange bias  and to the emergence of hybrid  collective modes that that combine FM and AF magnons.  The energy band of the heterobilayer show a type II band alignment, and feature spin-splitting of the states of the AF layer due to the breaking of the inversion symmetry.
\end{abstract}
\maketitle

\section{Introduction}
The observation of magnetic order in stand-alone monolayers\cite{wang2016,lee2016,Huang2017}
derived from van der Waals layered magnetic compounds has started a new
research area in 2D materials with potential for applications in spintronics.\cite{Gibertini2019}
Magnetic 2D crystals have also non-trivial physical properties such as topological
magnons,\cite{Chen2018,Costa2020} skyrmions,\cite{ding2019} or quantized
anomalous Hall effect.\cite{deng2020,Canonico2020} Moreover, the study of Van der Waals
heterostructures combining magnetic and non magnetic materials,\cite{Lyons2020} as 
well as different types of magnetic materials, creates a huge space of opportunities 
to create artificial structures with novel properties.\cite{Kezilebieke2020} This includes functional 
devices, such as spin filter tunnel junctions,\cite{Klein2018} as well as heterostructures 
where magnetic proximity effect promotes spin splitting in otherwise 
non-magnetic materials\cite{Lyons2020} as well as the 
emergence of topological superconductivity.\cite{Kezilebieke2020}

Here we choose to study heterobilayers made of two magnetic oxides like vanadium and iron oxychloride
(VOCl\cite{Glawion2009,Komarek2009,Ekholm2019,Yan2019} and
FeOCl\cite{Hwang2000,Bonacum2019}). 
Bulk FeOCl undergoes a paramagnetic to antiferromagnetic transition at 92$\pm$3 K, according to Mossbauer spectra\cite{Grant1971} with  antiferromagnetic order in the monolayers. 
In contrast, neutron scattering experiments\cite{Komarek2009,schonleber2009} shows that bulk  VOCl  
is formed by antiferromagnetically coupled monolayers with ferromagntic order,
very much like CrCl$_3$.\cite{mcguire2017} The Neel temperature of VOCl was
reported to be 80 K.\cite{schonleber2009,Komarek2009} Recently, the
ferrimagnetic phase of VOCl was reported at 150 K.\cite{Wang2021} Importantly, both
FeOCl\cite{Kim95} and VOCl\cite{Zhu2020b} are insulating.

Moreover, FeOCl has a unique catalytic structure\cite{Yang2013}
and VOCl has potential as electrode material in rechargeable 
lithium-ion batteries.\cite{Gao2015,Gao2016} Theoretical studies have
demonstrated that the bulk VOCl system has a strong magnetoelastic coupling
that might be used for development of magnetoelastic sensors and
actuators.\cite{Komarek2009} 

Apart of the current applications, our initial motivation to look into 
VOCl and FeOCl comes from the fact that degradation of samples exposed to air 
has been an practical obstacle in the experimental research on most 
magnetic 2D crystals. In principle, we expect that magnetic oxides are 
not affected by this problem, showing good air stability in recent
studies.\cite{Wang2021} In addition, most of the work so far 
has focused on ferromagnetic compounds, with the notable excepction 
of MPS$_3$ (M=Mn,Fe, Ni).\cite{Kim2019}

Another reason to study these compounds comes from their lattice structure. The 
magnetic atoms in FeOCl and VOCl form a square lattice, different from 
the honeycomb lattices of the chromium trihalides\cite{Soriano2020} or
FePS$_3$.\cite{wang2016,lee2016} This may bring new possibilities in the study of 
moire magnets\cite{hejazi2020} or in the study of proximity
effects.\cite{Hsieh2016}

In this work we address the electronic and magnetic properties of
van der Waals heterobilayers made of monolayers of FeOCl and VOCl. We 
explore emergent properties in the composite system that are missing in 
the constituent layers. We find that ferromagnetic (antiferromagnetic) order is preserved 
in the VOCl (FeOCl) monolayers, both on their own and forming part of a heterobilayer. We 
also report a type-II band alignment, useful for hosting long-lifetime
excitons,\cite{Torun2018} and a spin splitting of the otherwise 
spin degenerate bands of FeOCl in the range of 2 meV.

The manuscript is organized as follows. In Section II we briefly describe the
application of density functional theory togehter with the Hubbard correction to
obtain the main properties of monolayers and FeOCl/VOCl bilayers. In Section III we present our results 
for the monolayers and compare with the existing literature. In section IV we present the results
for the heterobilayer. In Section V we present our conclusions. 

\section{Methods}

We apply density functional theory (DFT),\cite{Burke2013} within
the generalized gradient approximation (GGA), Perdew-Burke-Ernzerhof (PBE) 
parameterization,\cite{Perdew1996} and the projector 
augmented-wave (PAW) method, including spin-orbit coupling, as implemented in the Vienna ab initio simulation
package (VASP).\cite{Joubert1999,Kresse1996} The electronic correlation of the
$d$ orbitals of Fe and V is computed with the Hubbard correction with values
of 5.3 and 2.0 eV for $d$ orbitals of Fe and V, 
respectively.\cite{Anisimov1991,Anisimov1997} In heterobilayers, the 
van der Waals interaction is accounted with the DFT-D3 method.\cite{Grimme2010}

The lattice constants and internal coordinates are optimized until the residual 
forces on each atom are less than 0.01 eV/$\AA$. We have optimized using
GGA+U. For all systems, a basis-set cutoff
energy of 520 eV is used, and the sampling of the Brillouin zone (BZ) is
a $10 \times 8 \times 1$ $\bf{k}$-grid. A vacuum layer of 20 $\AA$ is set in the
out-of-plane direction to avoid unphysical interactions between periodic replicas.

\begin{figure}[h]
\centering
\includegraphics[width=8cm]{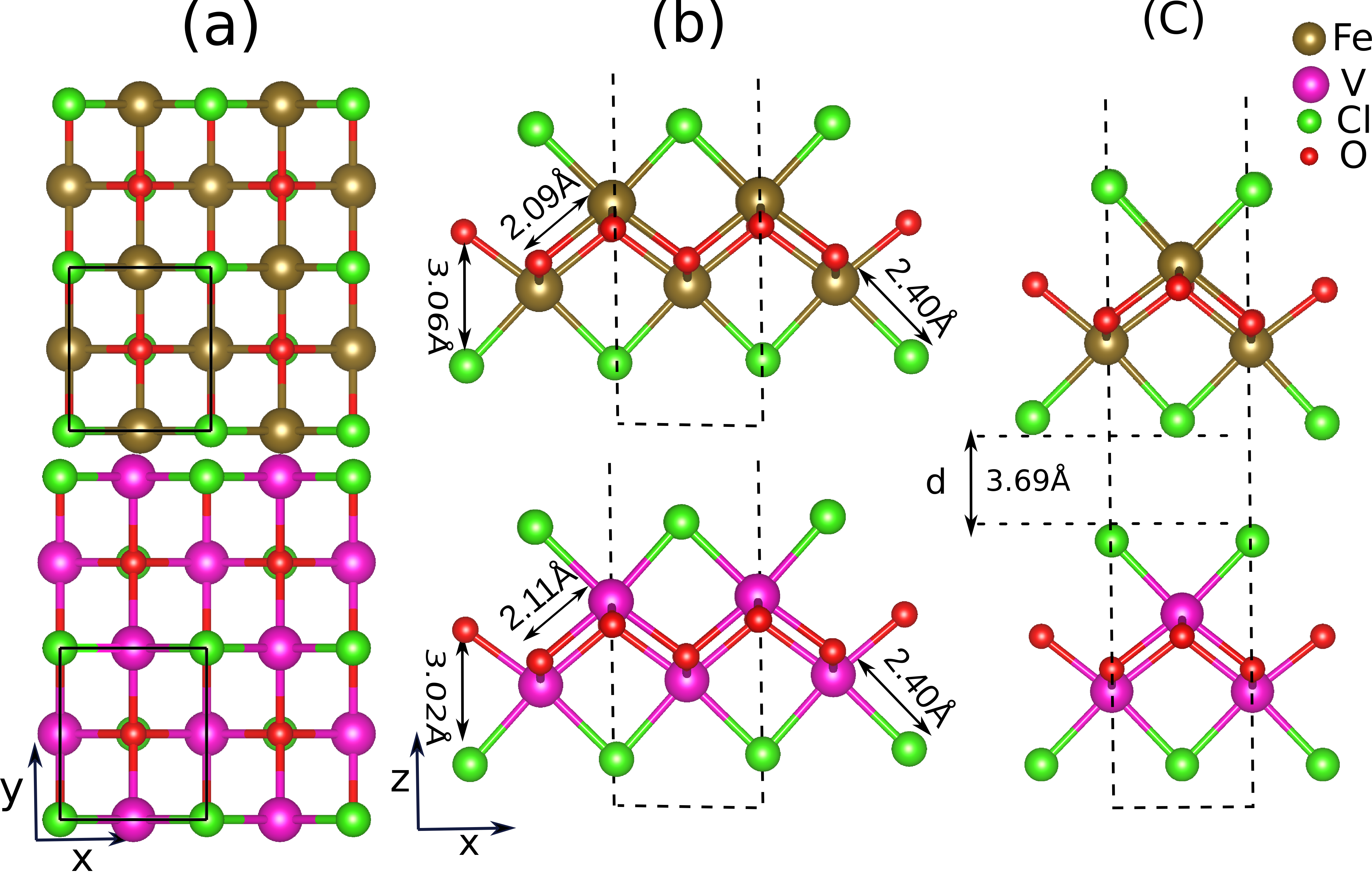}
\caption{(Color online) Top (a) and side (b) view of structure of FeOCl and VOCl 
monolayers, showing interatomic distances. (c) Side view of structure of heterobilayer 
FeOCl/VOCl. Dashed lines mark the  boundaries of the unit cells used in our DFT
calculations.}
\label{structure}
\end{figure}

The crystal structure of monolayer FeOCl and VOCl, together with the
FeOCl/VOCl heterobilayer is shown in Fig. \ref{structure}. The space group for
monolayers is $D_{2h}$ and the point symmetry is $C_{2v}$. In the case of the bilayer
the point symmetry is also $C_{2v}$. The monolayer in-plane lattice parameters are
$a=$ 3.281 \AA, $b= 3.836$ $\text{\AA}$ for FeOCl and $a=3.341$ $\text{\AA}$,
$b=$ 3.843 $\text{\AA}$ for VOCl.
In the case of the heterobilayer, the in-plane parameters are $a=$ 3.311
$\text{\AA}$, $b=$ 3.840 $\text{\AA}$, while the interlayer separation is $d=3.69$ \AA, as indicated in Fig.
\ref{structure}. The results of the structural optimization are in good agreement with previous 
works.\cite{Lind1970,Grant1971,haase1975vanadium}

\section{Electronic and magnetic properties of monolayers}

\subsection{Ab initio calculations}

Based on a simple analysis of the chemical valence of the atoms in
both compounds, we expect that  both Fe and V are in the $+3$ the oxidation state assuming that O and Cl have valence -2 and
-1, respectively. In the case of FeOCl, the external shell of Fe$^{3+}$ is 3$d^5$ and in the ionic 
limit relevant for an insulator, first Hund's rule predicts a spin
$S_{\text{Fe}}=5/2$, and thereby a magnetic moment  $M_{Fe}= 5\mu_B$. Our calculations
give $M_{\text{Fe}}= 4.8\mu_B$, not far from the naive ionic picture. We have also
found some residual magnetization of O and
Cl atoms. Analogously, for VOCl the external shell of V$^{3+}$ is 3$d^2$, and we
expect an atomic spin $S_{\text{V}}=1$, and a magnetic moment $M_{\text{V}}=2\mu_B$.  
In our calculations we found $M_{\text{V}}= 1.7\mu_B$, in line with the results for FeOCl.

From the ionic picture, V and Fe are not closed shell. Therefore, an important
question is whether the 3$d$ electrons form itinerant bands or, on the contrary,
a bandgap opens due to either crystal field splitting of the $3d$ levels or to 
correlation effects. From our DFT calculations we conclude that a bandgap only 
opens in FeOCl and VOCl monolayers when the Hubbard $U$ is included in the 
calculations, both when considering ferromagnetic (FM) and antiferromagnetic
(AF) order on the layers. As we discuss below, we find that FeOCl monolayers 
are AF whereas VOCl monolayers are FM.

Regarding the electronic bandgap, in both semiconductors the bandgap
is indirect. Nevertheless, in VOCl the conduction band (CB) minima is very close to the high
symmetry point $S$, in which yields the valence band (VB) maxima. Considering the relative
flat bands at VB and CB edges, the absorption/emission should be still
comparable to that of direct 2D materials.

\begin{figure}[h]
\centering
\includegraphics[width=8 cm]{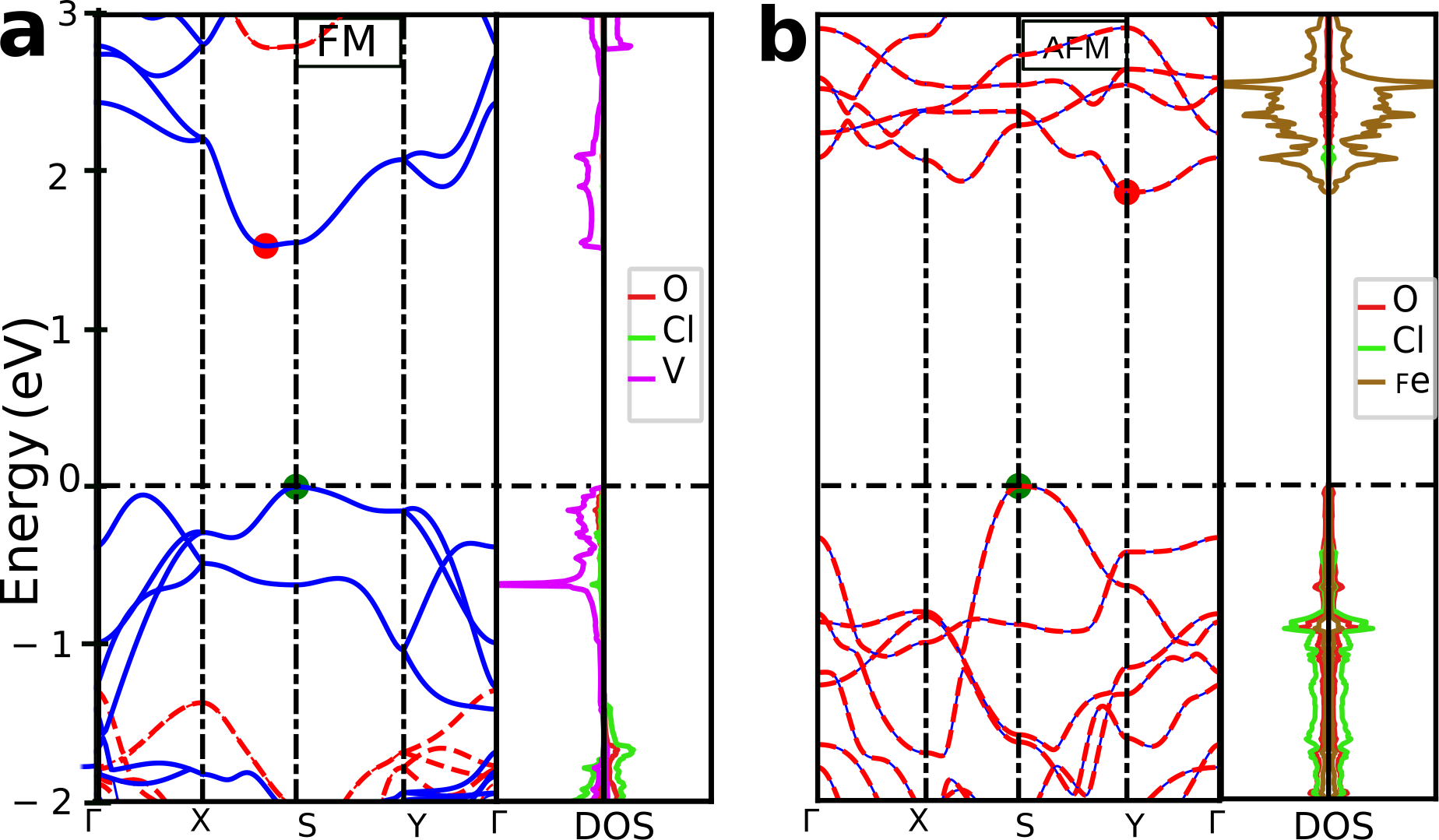}
\caption{Band structure of monolayer FeClO-AF and VOCl-FM calculated within
PBE$+U$. The projected DOS is represented using the atomic color 
code of Fig. \ref{structure}.}
\label{monolayer-af}
\end{figure}

Figure \ref{monolayer-af} shows the band structures of FeOCl and VOCl monolayers
computed with PBE$+U$ in the AF and FM phase, respectively, together with the projected DOS. In the case of FeOCl the
hybridization of Fe $d$ orbitals with the Cl and O atomic orbitals is very strong, as
shown in the projected DOS of the valence band. Note that he conduction band is purely composed by iron.
Moreover, the bands of FeOCl are doubly degenerate as expected on an antiferromagnet with
an inversion center. In the case of VOCl the situation is pretty different, for two reasons: first, 
hybridization with O and Cl in both valence and conduction bands is very weak. Second, 
both valence and conduction band are spin polarized, with spin projection along the 
majority spin direction. We note that the top of the valence band of VOCl has
a two-fold orbital degeneracy along the $SY$ and $SX$ directions.

In order to determine both the type of magnetic order and the magnetic
anisotropy, we have carried out PBE$+U$ calculations considering both FM and AF configurations 
with spin pointing along four directions [100], [010], [001], and [110]. A summary of the ground 
state energies for different configurations is shown in Table \ref{mae}. Our
PBE$+U$ calculations show that the AF and
FM spin configurations are the stable phases for FeOCl and VOCl monolayer,
respectively, in agreement with previous
works\cite{Glawion2009,wang2020discovery,Mounet2018}. 
In both compounds the ground states magnetization is lying
along the [010] axis, the spin points along the V-O/Fe-O bond direction, projected along the plane of the layer.

\subsection{Spin model}

Given the insulating nature of the monolayers, we propose a spin model to describe the 
DFT calculations. The model has 3 types of terms, first-neighbour isotropic Heisenberg 
exchange, first-neighbour anisotropic exchange (ae) and single ion anisotropy
(sia):

\begin{equation}
H= H_{\rm Heis} + H_{\rm ae} + H_{\rm sia},
\end{equation}
where:
\begin{equation}
H_{\rm Heis}= \sum_{i,i'}\frac{J}{2}\vec{S}(i)\cdot\vec{S}(i') 
\end{equation},
\begin{equation}
H_{\rm ae}=
\frac{J_y}{2}\sum_{i,i'} S_y(i)S_y (i')+
\frac{J_z}{2}\sum_{i,i'} S_z(i)S_z (i'),
\end{equation},
and:
\begin{equation}
H_{\rm sia}=\sum_i\left( E( S_x(i)^2-S_z(i)^2)  - D S_y(i)^2 \right).
\end{equation}

In the sums above $i$ runs over all the lattice sites of a square 
lattice and $i'$ runs over the 4 first neighbours of $i$. The first term 
accounts for first-neighbour isotropic exchange and the third term accounts 
for single ion anisotropy that reflects the symmetry of the crystal field 
of Fe and V atoms. In total, we have 5 parameters $J,E,D,J_y,J_z$. We estimate 
them using 6 DFT calculations (see Appendix) and then we test the model with
the remaining two. The results of the fitting gives an error with respect the
DFT calculations of 7.95 \% and 6.09 \% for FeOCl and VOCl, respectively, which
is rather satisfactory. 

In both cases the single ion anisotropies $D,E$ are larger than the anisotropic exchange 
constants $J_y,J_z$.  However, this difference is much smaller  in the case of FeOCl. The smaller single ion anisotropy  
values in this case are probably due to the fact that the $d$ shell is half full for Fe, so that the orbital momentum is 0, even without crystal field quenching. 
The predominance of single ion anisotropies, in both cases, is to be expected on 
account of the smaller atomic weight of the ligands, and thereby smaller spin-orbit coupling. We also note that the proposed model 
has two main limitations: it only considers first neighbour exchange and it completely 
ignores magneto-elastic interactions, that are known to be important for VOCl.\cite{Komarek2009}

\section{Electronic and magnetic properties heterobilayers}

We now address the electronic and magnetic properties of the
VOCl/FeCl heterobilayer. We assume the bulk stacking configuration
for the heterobilayer, as shown in Figure \ref{structure}c. As in the case of monolayers, we 
have explored several magnetic states. We consistently found that both 
the magnetic order type and the easy axis of the monolayers is preserved 
in the heterostructure. This indicates that interlayer exchange is weaker 
than both the intralayer exchange and the magnetic anisotropy.

The isolated FeOCl has two equivalent N\'{e}el states, that are related by time-reversal symmetry 
(inversion of the magnetization direction). Because of the buckling of the Fe atoms, interlayer exchange 
breaks this degeneracy, as one of the Fe magnetic sublattices is closer to the vanadium atoms than the other. 
We refer to these two phases as parallel (p) and anti-parallel (ap), for the FM and AF relative spin 
orientation of the two closest V-Fe atoms.  Our calculations indicate that interlayer exchange 
is ferromagnetic. The difference of energy difference per unit cell, $\Delta E_{12}$ between the 
two states is related to the interatomic exchange as:

\begin{equation}
  \Delta E_{12}= j S_{\text{V}} S_{\text{Fe}},
\end{equation}
where $j$ is the interlayer exchange coupling. We are ignoring here the small anisotropy 
of this quantity, which is already small. Indeed, we have $\Delta E_{12} = 0.14
$ meV from which  we infer $j\simeq 0.056$ meV. This interlayer exchange should lead to 
exchange bias of the hysteresys cycle of the VOCl 
monolayer as well as the splitting of the otherwise degenerate AF magnons of the FeOCl and very likely their  hybridization, at finite $q$ with the FM magnons of the VOCl bilayer. This hybridization is expected because the  AF magnons  have a larger $q=0$ magnon gap than the FM magnons but a smaller dispersion, on account of their linear dispersion. At the crossing point, interlayer exchange should lead to hybridization gap driven by the interlayer exchange. This will be the subject of future work.

\begin{table}
	\resizebox{0.5\textwidth}{!}{
		\begin{tabular}{llcccc}
			\hline 
			\hline 
      2D Crystal & GS  & $E_{[100]}$ & $E_{[010]}$ & $E_{[001]}$ & $E_{[110]}$ \\
			\hline                                                                           
			FeOCl      & FM  & 54.635      &  54.396     & 54.797     & 54.518       \\
			           & AF  &  0.201      & \textbf{0}  &  0.275     & 0.102        \\
			\hline                                                                           
			\hline                                                                           
			VOCl       & FM  & 0.443       & \textbf{0}  & 1.326      & 0.236        \\
		       	     & AF  & 24.539      & 24.071      & 25.379     & 24.334       \\
			\hline 
			\hline
			VOCl/FeOCl & AF(Fe)/FM(V)-p  & 0.585  & \textbf{0}  &  1.614  &  0.340   \\
					   		 & AF(Fe)/FM(V)-ap & 0.733  & 0.140       &  1.750  &  0.482   \\
			\hline 
			\hline
	\end{tabular}}
\caption{Magnetocrystalline anisotropy energies $E_{[100]} - E_{[010]}$,
  $E_{[001]} -E_{[010]}$, and $E_{[110]} -E_{[010]}$(meV/ unit cell). The
  reference is the total energy of the 101 direction, $E_{[010]}$, 
which in all cases is found to be the ground state. For the heterobilayer AF(Fe)/FM(V)-p stands for parallel
  spin polarization of neighbor Fe and V atoms and AF(Fe)/FM(V)-ap for
  anti-parallel spin polarization.} 
	\label{mae}
\end{table}

\begin{table}
\begin{tabular}{ccc}
	\hline 
	\hline 
         &  FeOCl    & VOCl     \\
	\hline 
  $S$    &  $5/2$    &  1       \\
  $J$    &  2.177 meV  & -6.024 meV  \\
  $J_y$  & -1.6 $\mu$eV   &  6.2 $\mu$eV  \\ 
  $J_z$  &  3.5 $\mu$eV   &  10.7 $\mu$eV  \\
  $D$    &  5.1 $\mu$eV   &  861.1 $\mu$eV  \\
  $E$    & -9.5 $\mu$eV   & -430.7 $\mu$eV \\
  $\epsilon_{110}-\epsilon_{010}$ & 0.1196  meV & 0.2089 meV \\
  $\Delta$(\%)  &  7.95   & 6.09 \\ 
	\hline 
	\hline 
\end{tabular}
\caption{Parameters of the spin model of monolayer FeOCl and VOCl, expressed in
  meV. The energy difference $\epsilon_{110}-\epsilon_{010}$ obtained from the
  model is the AF case for FeOCl and the FM case for VOCl. The error is
  $\Delta=|(\Delta_{\text{model}}-\Delta_{\text{DFT}})/(\Delta_{\text{model}}+\Delta_{\text{DFT}})|*100$.
 }
\label{table-model}
\end{table}

\begin{figure}[h!]
	\centering
	\includegraphics[width=8cm]{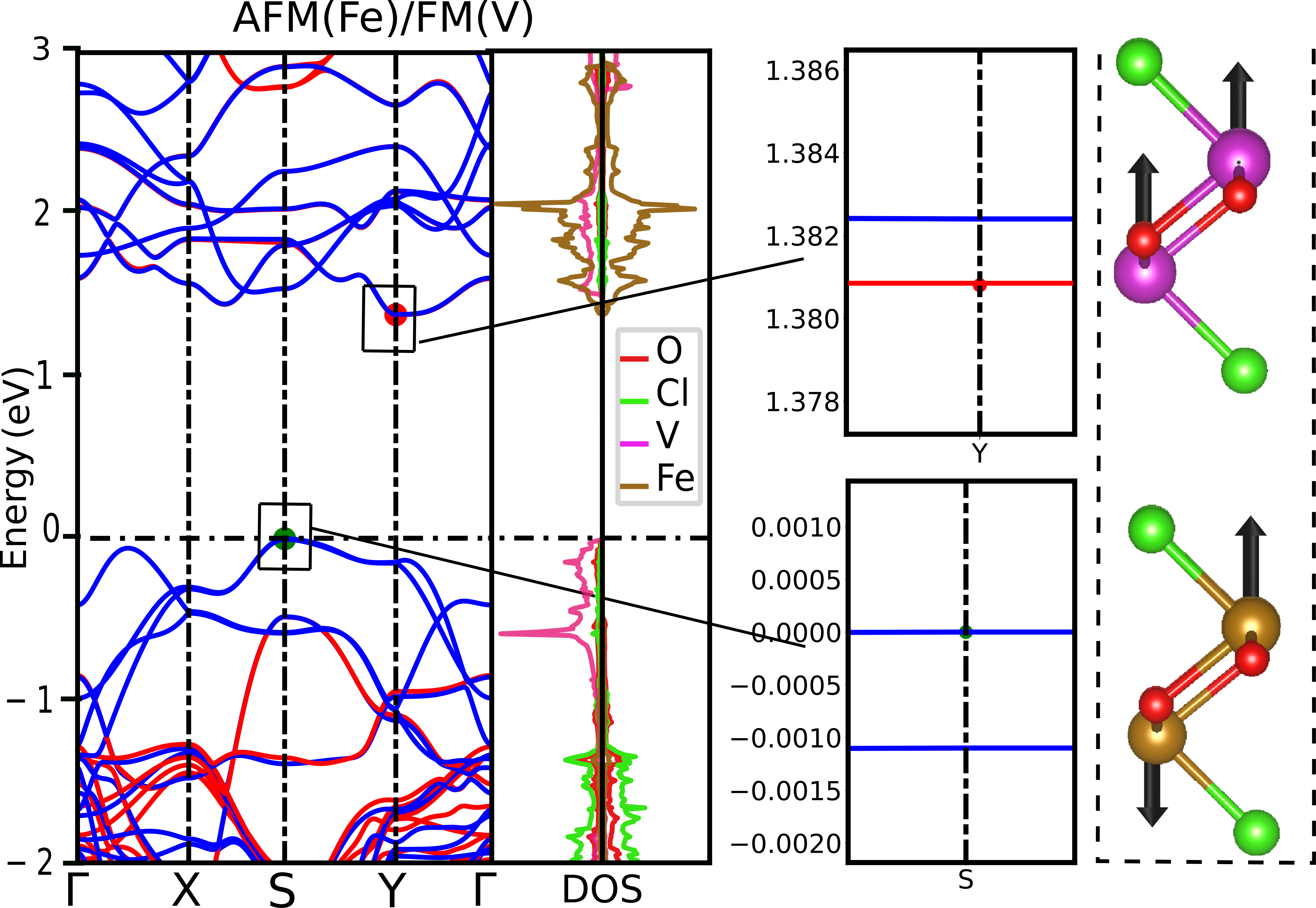}
	\caption{Band structure of Heterobilayer FeOCl(AFM)/VOCl(FM).Insets show the detail of VBM and CBM and
		the spin splitting.}
	\label{bilayer}
\end{figure}

We now discuss the electronic properties of the hetero-bilayer. Figure \ref{bilayer} shows 
the band structure and the projected DOS of the most
stable magnetic state of the heterobilayer. The bilayer exhibits an indirect bandgap of 1.38
eV. From the projected DOS we determine that the band alignment is type II, 
similar to the case of heterostructures of transition metal dichalcogenides.\cite{kosmider13,Torun2018} The top of the
VB is mainly composed by V orbitals while the bottom of the CB 
is dominated by the Fe wave functions. Interestingly, we find that both band
extrema feature splittings missing in the monolayer. In the case of the
CB, we find a spin-splitting of the Fe bands. This can arise
either from the interlayer  spin exchange or from the breaking of inversion symmetry
of the heterobilayer. The fact that the sign of the splitting is the same for
the two interlayer alignments, parallel and antiparallel, discussed above is a strong suggestion
that the origin of the splitting is due to the absence of inversion symmetry.\cite{soriano12} On the 
other hand, the splitting observed in the top of 
the VB occurs between two states with the same spin and is related to
reduced symmetry of the heterobilayer.

\section{Conclusions}

We have characterized the electronic and magnetic properties of VOCl/FeOCl
heterobilayers using \textit{ab initio} methods, including the electronic correlation
effects within the framework of the Hubbard method.
 Consistent with experimental work, we find that VOCl (FeOCl) 
monolayers are ferromagnetic (antiferromagnetic) insulators with an easy axis along the [010] direction. We find a hybrid 
magnetic order for VOCl/FeOCl heterobilayers, that combines the ferromagnetism
of the VOCl monolayer and the antiferromagnetic order in the FeOCl layer.
Interlayer exchange is found to be ferromagnetic and should result in a bias of
the hysteresis cycle of the VOCl monolayer. Moreover, the reduced symmetry of the heterobilayer
causes a spin splitting of both
conduction and valence band edges of $\simeq 2$ meV. The hetero-bilayer has  a 
type-II band alignment, with the conduction/valence band localized in 
the Fe(V) layer. Interestingly, combined with the spin splitting of the valence band 
states, interlayer excitons in this system would have a well-defined spin
polarization. Therefore, the magnetic
properties can be combined with a long lifetime for 
the excitons. We also foresee the emergence of hybrid magnon modes combining the AF
and FM magnons of both layers.

\section{Acknowledgments}
This work has been supported by the Algerian Ministry of High
Education and Scientific Research under the PNE programme. F. M. thanks the hospitality of the Departamento de F\'isica Aplicada at the Universidad de Alicante and we also thank W. Aggoune for his outstanding help and valuable advice. A. M.-S. acknowledges the Ram\'on y Cajal programme (grant RYC2018-024024-I; MINECO, Spain)
and the Marie-Curie-COFUND program Nano TRAIN For Growth II (Grant Agreement 713640). Ab 
initio simulations were performed performed on the Tirant III cluster of the Servei 
d‘Informática of the University of Valencia. J. F.-R. acknowledges financial support from 
 Generalitat Valenciana funding 
Prometeo2017/139,
FEDER/Junta de Andaluc\'ia CTEICU, grant 
Proyecto PY18-4834 and
MINECO-Spain (Grant No.PID2019-109539GB). 

\appendix

\section{Determination of parameters of the spin model}

The parameters $J$, $J_y$,$J_z$ $D$ and $E$ are calculated from the classical
energies per atom:

\begin{eqnarray}
&&\epsilon_{\eta}(\vec{n})\equiv\frac{E_{\eta=\pm 1}(\vec{n})}{N_{\rm at}} = S^2\frac{\eta z (J+ J_y n_y^2 + J_z n_z^2) }{2} +\nonumber\\
&&+S^2 \left(E(n_x^2-n_z^2) - Dn_y^2 \right),
\end{eqnarray}

where $\eta=+1 (-1)$ stands for FM (AF) configurations,  $S=1, 5/2$ is the spin
offor V and Fe, respectively, $z=4$ is the number of first neighbours, and
$\vec{n}=(n_x,n_y,n_z)$ is the unitary vector that marks the different
orientations of the magnetic moments. The single ion magnetic anisotropy does
not depend on $\eta$. We obtain 6 configurations (FM and AF) for the
three different orientations of the magnetic moments along the crystal axis:

\begin{eqnarray}
&&\epsilon_{\eta}(\vec{n_{x}})\equiv\frac{E_{\eta=\pm 1}(\vec{n_{x}})}{N_{\rm
  at}} = S^2\frac{\eta z (J) }{2} \nonumber+S^2 ( E(n_x^2)),
\end{eqnarray}

\begin{eqnarray}
&&\epsilon_{\eta}(\vec{n_{y}})\equiv\frac{E_{\eta=\pm 1}(\vec{n_{y}})}{N_{\rm
  at}} = S^2\frac{\eta z (J+ J_y n_y^2) }{2} +\nonumber S^2(-Dn_y^2),
\end{eqnarray}

\begin{eqnarray}
&&\epsilon_{\eta}(\vec{n_{z}})\equiv\frac{E_{\eta=\pm 1}(\vec{n_{z}})}{N_{\rm
  at}} = S^2\frac{\eta z (J+ J_z n_z^2) }{2} \nonumber+S^2 E(-n_z^2).
\end{eqnarray}

In order to derive $J$ we can use:

\begin{equation}
\epsilon_{FM}(100)-\epsilon_{AF}(100)= S^2z J .
\end{equation}

We now  can obtain $J_y$ and $J_z$ from
\begin{equation}
\epsilon_{FM}(010)-\epsilon_{AF}(010)=
S^2z J+ S^2 z J_y 
\end{equation}
and
\begin{equation}
\epsilon_{FM}(001)-\epsilon_{AF}(001)=
S^2z J+ S^2 z J_z 
\end{equation}

In order to derive the single ion anisotropies, $E$ and $D$ we now compare FM energies along different directions:

\begin{equation}
\epsilon_{FM}(100)-\epsilon_{FM}(010)=DS^2+ES^2+\frac{J_y zS^2}{2}
\end{equation}
and
\begin{equation}
\epsilon_{FM}(001)-\epsilon_{FM}(010)= DS^2-ES^2+\frac{J_y zS^2}{2}-\frac{J_z zS^2}{2}
\end{equation}

with two equations and two unknowns. Table \ref{table-model} summarizes the
parameters of monolayer FeOCl and VOCl obtained from the DFT values of Table
\ref{mae}. 


%
\end{document}